\documentclass[journal=jctcce,manuscript=article,layout=twocolumn]{achemso}
\usepackage{etoolbox}
\usepackage{multirow}
\usepackage{tikz}
\usepackage{subcaption}
\usepackage{amsmath}
\usepackage{color, soul}

\title{A set of moment tensor potentials for zirconium with increasing complexity}
\author{Yu Luo}
\affiliation{Department of Mechanical and Materials Engineering, Queen’s University, Kingston, Ontario, K7L2N8, Canada}
\author{Jason A. Meziere}
\affiliation{Department of Physics, Brigham Young University, Provo UT 84602 USA}
\author{German D. Samolyuk}
\affiliation{Materials Science and Technology Division, ORNL, Oak Ridge TN 37831-6138 USA}
\author{Gus L. W. Hart}%
\affiliation{Department of Physics, Brigham Young University, Provo UT 84602 USA}%
\author{Mark R Daymond}
\affiliation{Department of Mechanical and Materials Engineering, Queen’s University, Kingston, Ontario, K7L2N8, Canada}
\author{Laurent Karim Béland}
\affiliation{Department of Mechanical and Materials Engineering, Queen’s University, Kingston, Ontario, K7L2N8, Canada}
\email{laurent.beland@queensu.ca}

\begin{document}

\onecolumn

\begin{abstract}
Machine learning force fields (MLFFs) are an increasingly popular choice for atomistic simulations due to their high fidelity and improvable nature. Here, we propose a hybrid small-cell approach that combines attributes of both offline and active learning to systematically expand a quantum mechanical (QM) database while constructing MLFFs with increasing model complexity. Our MLFFs employ the moment tensor potential formalism. During this process, we quantitatively assessed structural properties, elastic properties, dimer potential energies, melting temperatures, phase stability, point defect formation energies, point defect migration energies, free surface energies, and generalized stacking fault (GSF) energies of Zr as predicted by our MLFFs. Unsurprisingly, model complexity has a positive correlation with prediction accuracy. We also find that the MLFFs wee able to predict the properties of out-of-sample configurations without directly including these specific configurations in the training dataset. Additionally, we generated 100 MLFFs of high complexity (1513 parameters each) that reached different local optima during training. Their predictions cluster around the benchmark DFT values, but subtle physical features such as the location of local minima on the GSFE surface are washed out by statistical noise. 
\end{abstract}


\twocolumn

\section{Introduction}
Zr and Ti have an unfilled d-electron band\cite{bacon2002atomic, bakonyi1993electronic}. This leads to more directional bonding in these hcp materials than in most metals. In turn, this particular bonding leads to key properties, including an hcp to bcc martensitic transition, a low vacancy migration barrier (considering their high melting point), an increase in electronic conductivity as temperature increases, and specific features of their generalized stacking fault energy surfaces\cite{aguayo2002elastic,grad2000electronic,roy2022vacancy,zhou2017ab,vohra1979electronic,yu2015high,jamieson1963crystal,benoit2012density,chen2022mesoscale}. 

As shown in a study by Oliver \textit{et al}~\cite{nicholls2023transferability}, current interatomic force fields fail to capture many of these key physical properties of Zr which imposes important compromises on their users. Both embedded atom method (EAM) models and MLFFs significantly underestimate zirconium's melting point, with the exception of an EAM model specifically designed to model its melting point~\cite{mendelev2007development}; the latter which captures the correct melting point drastically overestimate vacancy migration barriers. In the context of modeling radiation-induced defects, to the best of our knowledge, no existing force fields can accurately capture the relative jump barriers of vacancies and self-interstitial atoms (SIA) along basal and non-basal directions. Vérité \textit{et al}~\cite{verite2007anisotropy} used density functional theory (DFT) calculations which displayed much lower vacancy migration barriers compared to EAM models. Later, DFT calculations by Samolyuk \textit{et al}~\cite{samolyuk2014analysis} indicated the anisotropy in jump barrier between basal and prismatic SIA jump barriers is in fact smaller than that of vacancies', which current Zr potentials do not capture. 

Another limitation of current interatomic force fields for zirconium pertains to the generalized stacking fault energy (GSFE) surfaces. For instance, Clouet \textit{et al}~\cite{clouet2012screw} identified spurious local minima at the basal GSFE surface along the [1$\overline{1}$00] direction as predicted by EAM models\cite{mendelev2007development}. This spurious local minimum in the basal plane can lead to unexpected screw dislocation dissociation in large-scale dislocation modeling. Additionally, a GSFE local minimum at 1/6[1$\overline{2}$10] is absent or shifted along the [0001] direction, while screw dislocation dissociation into two partials in the prismatic plane of zirconium is predicted by DFT calculations. These artifacts on the GFSE surface are thought to be inherent to central-force potentials that do not explicitly consider the angular interactions terms~\cite{bacon2002atomic}. 

MLFFs open the door to addressing these issues\cite{bartok2010gaussian,behler2007generalized, botu2017machine}. However, some recent attempts have not been fully successful in capturing all of zirconium's critical properties. Qian \textit{et al}~\cite{qian2018temperature} created Gaussian approximation potential\cite{bartok2010gaussian,bartok2013representing} models for investigating temperature effect on phonon dispersions in Zr, but these models perform poorly in out-of-sample scenarios, such as when considering Zr dimers in vacuum \cite{nicholls2023transferability}. Zong \textit{et al}~\cite{zong2019hcp} adopted a kernel ridge regression model to capture the martensitic phase transformations in Zr, but this resulted in a low melting point\cite{nicholls2023transferability}. Nitol \textit{et al}~\cite{nitol2022machine} modeled the transformations among $\alpha$, $\beta$, and $\omega$ phases of Zr and Ti using artificial neural networks as regression models\cite{behler2007generalized}, and Liyanage \textit{et al.} \cite{liyanage2022machine} developed an artificial neural network potential for hcp Zr that focused on extended defect properties. These sophisticated neural network architectures can possess up to tens of thousands of parameters, fitted on over ten thousand DFT configurations. While these neural network potentials can capture many physical properties of Zr with high-fidelity (albeit performance out-of-sample still raises questions), their training and use involves a large computational cost. These examples demonstrate the limitations of existing MLFFs of Zr. More effort is needed to address these issues and develop MLFFs that can capture critical physical features of Zr in a more data-efficient way. 

Note that the referenced works above employ so-called offline approaches, which may explain the limitations; the developer-scientist has to pick all possible configurations and select the right hyperparameters. Collecting all necessary configurations is challenging because it is computationally demanding in QM calculations and difficult to ensure that configurations are representative and diverse enough for the training data set. Segmenting the dataset into training and validation sets is a good first step but is not a guarantee of robustness. To address this issue, Karabin \textit{et al}~\cite{karabin2020entropy} proposed an entropy-maximization approach to sample diverse local atomic environments for the generation of training datasets in force fields. This promising approach leads to models that perform well out-of-sample. However, it tends to lead to fitting processes with relatively large training errors.

Active learning approaches, where learning algorithms adjust the configuration generation process as part of the training progress, address some of these issues\cite{li2015molecular,podryabinkin2017active}. For example, Grigorev \textit{et al}~\cite{grigorev2023calculation} deployed a hybrid method combining QM calculations and predictions from linear machine learning force fields and validated its soundness over helium and vacancy segregation to dislocations. However, in many scenarios, running active learning algorithms is not feasible or desirable due to factors such as time constraints while continuously updating models, identifying the most efficient QM/ML geometry, computational resource limitations, and difficulty in conducting QM calculations for sophisticated systems to extend dataset. 

In this article, we first propose a hybrid approach that combines some of the advantages of both offline and active learning to systematically expand the QM training data set for creating MLFFs. Secondly, we present our physical validations for resulting force field models, which include crystalline structures, melting behaviors, surface structures, point defects, and extended defects. Thirdly, we conduct 100 trials to develop MLFFs to investigate their sensitivity to different random seeds before model optimization, showcasing the effectiveness and limitations of our approach. Fourthly, we propose our optimal MLFFs for Zr based on a decision tree. Finally, we draw conclusions.

\section{Computational methods} 
In this work, we employ the moment tensor potentials (MTP) framework.\cite{shapeev2016moment} In a previous comparative benchmark study, MTPs displayed a good trade-off between computational efficiency and accuracy\cite{zuo2020performance}. They can be adapted to different levels of complexity depending on the needs. Fitting MTPs comprises a mathematically well-defined procedure, which assumes total energy $E_\mathrm{mtp}$ in a system as a linear combination of energy contributions associated to a set of structural descriptors $V$ for each atom in the system.
\newline
\begin{equation}
E_\mathrm{mtp} = \sum_{i = 1} V(n_{i}) 
\end{equation}
\newline
where $n_{i}$ stands for neighborhood atoms for each atom. Atomic configurations can be mapped by a set of structural descriptors describing the local environment around each atom regarding neighborhood atom coordinates contained in a sphere of cut-off centered on the central atom. Theoretically, any continuous function describing local environments can be approximated by linear combinations of polynomial basis functions ($B$). 
\newline
\begin{equation}
V(n_{i}) = \sum CB(n_{i}) 
\end{equation}
\newline
where $C$ denotes slope coefficients and $B(n_{i})$ results from the contraction of the moment tensor that describes the local environment of each atom. Further details are presented here\cite{shapeev2016moment}. 
\newline
\begin{figure}[htbp]
  \hspace*{-0.03\linewidth}
  \begin{subfigure}[b]{0.52\textwidth}
    \includegraphics[width=\textwidth]{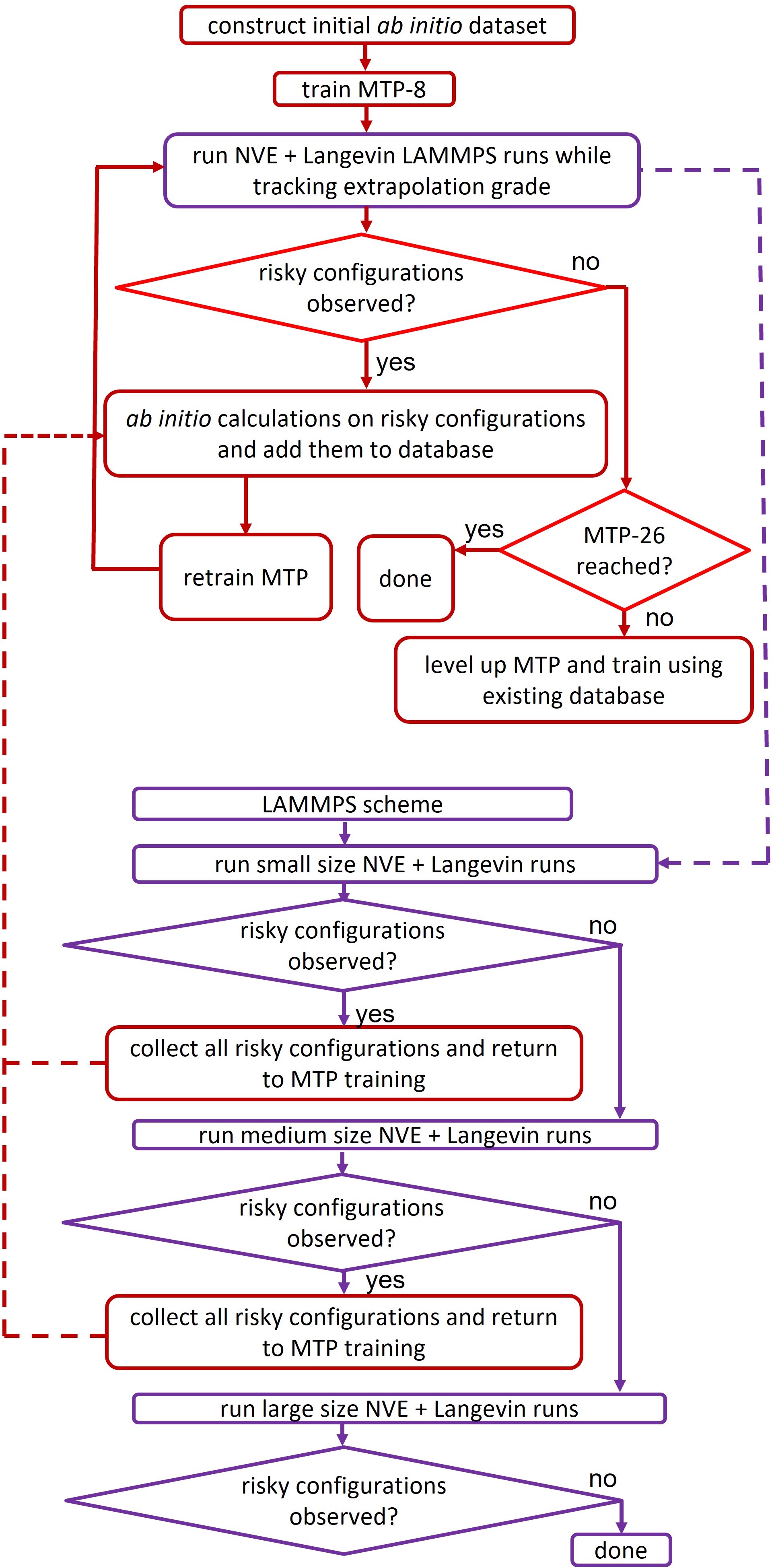}
  \end{subfigure}
  \caption{A flowchart illustrating our hybrid scheme combining offline and active learning to systematically expand the \textit{ab initio} dataset while simultaneously increasing the complexity of the MTP model. The top section of the flowchart (red) indicates how the complexity of the model increases in tandem with the size of the dataset. The bottom section of the flowchart (purple) outlines the LAMMPS\cite{plimpton1995fast} scheme to collect risky configurations. The MLIP-2 package is employed to fit the MTPs  \cite{novikov2020mlip}. MTP-8 and MTP-26 represent untrained MTP models with parameters of 26 and 1513, respectively.} 
  \label{fig:1}
\end{figure}
The systematic construction of an MTP is involves a highly non-linear mapping to connect atomic coordinates with their corresponding energies, forces, and stress tensors, typically done using supervised learning. Conducting \textit{ab initio} calculations for all possible atomic coordinates is computationally expensive and not realistic. Here, necessary methodologies composed in Figure \ref{fig:1} are presented to mitigate this issue. First, we constructed our initial structural dataset comprising a total of 1224 small-cell \textit{ab initio} structures and trained it with a lightweight, 26-parameter MTP (MTP-8). Further details related to the initial structural dataset can be found in the Supplementary Information. 

Inherent uncertainty and instability of atomic systems are expected in MD simulations because of unseen local environments induced by continuous atomic motion. To address this concern, MD simulations were run using the Large-scale Atomic/Molecular Massively Parallel Simulator (LAMMPS) ~\cite{plimpton1995fast} in the NVT ensemble using a Langevin thermostat at 300K, 800K, 1300K, 1800K, and 2300K, respectively. Throughout this process, an extrapolation grade criterion \cite{novikov2020mlip} based on D-optimality to estimate to which extent adding the configuration to the training set would increase its hypervolume in configurational space, which correlates with configuration prediction error without incorporating prior knowledge of \textit{ab initio} results. If risky configurations are observed, which are defined to have an extrapolation grade threshold larger than 2.1, we collect existing risky configurations found during these MD runs and remove any configurations that are structurally similar to each other. 

QM calculations are performed on the remaining configurations and their QM results are updated to the current dataset. We then retrain the MTP model using this updated dataset. The process described above is repeated until no risky configurations are observed in MD simulations. Then, once no risky configurations are observed, MTP model complexity is increased and the updated dataset of the current round serves as the initial dataset for the next round. This process is repeated until the desired level of model complexity is reached, which allows for continuous improvement of MTPs in a data-efficient pattern. 

The initial dataset leads to sparse regions in high-dimensional space due to the curse of dimensionality\cite{donoho2000high}. The process as outlined in Figure \ref{fig:1} addresses this concern by consistently introducing data-efficient (small-cell) risky configurations to the initial data set. 

MD simulations were performed with the open-source code LAMMPS ~\cite{plimpton1995fast}. Periodic boundary conditions are applied to all our atomic configurations. All runs are 100 ps in duration, with a time step of 1 fs. As shown in Table \ref{table:1}, multiple crystal structures, point defects, line defects, and planar defects are investigated using the LAMMPS scheme illustrated in Figure \ref{fig:1}.

This process starts by running LAMMPS on small-size atomic systems, as suggested by Mezière \textit{et al.} for accelerating force field fitting\cite{meziere2023accelerating}. These correspond to systems with fewer than 14 atoms. As shown in Table \ref{table:1}, 1552 small-cell initial configurations were considered. If risky configurations are observed during the small-cell MD runs, the process is stopped, the risky configurations are collected, DFT calculations are performed, the MTP is retrained, and all 1552 small-cell MD runs are performed once again. This iterative process continues up until no small-cell risky configurations are observed. Afterwards, we perform MD runs using 1834 medium-sized (involving 15–38 atoms) different initial configurations. These are needed to ensure interatomic interactions associated to planar defects are considered. We then run the same iterative risky configuration collection process as described above for small-cell configurations. Changing system size can affect the physical properties of defects and many interesting phenomena only appear in a large system. Therefore, the last phase of the training procedure involves performing large-scale (more than 38 atom; up to 512 atoms) MD simulations, using the same iterative scheme as for small-cell and medium-sized cell systems.

Note the majority of risky configurations collected were small-cell, and a few medium-size configurations were observed. No risky large-sized configurations--which are problematic for DFT calculations--were discovered by our LAMMPS scheme.

\begin{table*}[]
\center
\caption{Categorized list of atomic configurations used as inputs for the LAMMPS scheme. Values outside of parentheses correspond to the number of atomic configurations, while values in parentheses describe the number of atoms involved in each configuration.}
\resizebox{2\columnwidth}{!}{
\begin{tabular}{c|ccc}
\hline
                                      & \textbf{\begin{tabular}[c]{@{}c@{}}Small Size\\  ( \textless 15 atoms)\end{tabular}} & \textbf{\begin{tabular}[c]{@{}c@{}}Medium Size\\  ( 15 - 38 atoms)\end{tabular}} & \textbf{\begin{tabular}[c]{@{}c@{}}Large Size\\  ( \textgreater 38 atoms)\end{tabular}} \\ \hline
\textbf{Deformed HCP}                 & 39 (4)                                                                                & None                                                                             & None                                                                                    \\
\textbf{Deformed BCC}                 & 32 (2)                                                                                & None                                                                             & None                                                                                    \\
\textbf{Deformed FCC}                 & 98 (4)                                                                                & None                                                                             & None                                                                                    \\
\textbf{SIA-related structures}       & 8 (5), 8 (9)                                                                           & 8 (17), 8 (33)                                                                     & 8 (65), 8 (129)                                                                           \\
\textbf{Vacancy-related structures}   & 2 (3), 2 (7)                                                                           & 2 (15), 2 (31)                                                                     & 2 (63), 2 (127)                                                                           \\
\textbf{Divacancy-related structures} & 1 (6), 1 (14)                                                                          & 1 (30)                                                                     & 1 (62), 1 (126), 1 (254)                                                                          \\
\textbf{Edge Dislocations}            & None                                                                                 & None                                                                             & 2 (512)                                                                                  \\
\textbf{Free Surfaces}                & 8 (12)                                                                                & \begin{tabular}[c]{@{}c@{}}8 (16), 8 (20), 11 (24), \\ 11 (28), 11 (32)\end{tabular}  & 3 (40), 3 (44)                                                                            \\
\textbf{Generalized Stacking Faults}  & 882 (2), 441 (12)                                                                      & 882 (16), 882 (24)                                                                 & 882 (64)                                                                                 \\ \hline
\end{tabular}}
\label{table:1}
\end{table*}


To perform the DFT calculations, the generalized gradient approximation~\cite{perdew1996generalized} and the exchange-correlation functional parameterized by Perdew-Burk-Ernzerhof \cite{perdew1996generalized} were applied using the Quantum ESPRESSO software package~\cite{giannozzi2009quantum} with the pseudo-potential (Zr.pbe-nsp-van.UPF). We used a basis of plane waves with a kinetic energy of wave-functions cut-off of 104 eV. An energy convergence threshold of $10^{-4}$ eV and a force convergence threshold of $10^{-3}$ atomic units are applied. The number of k-points were varied depending on the size of the super-cell.

\section{Results and discussion}
\begin{figure}[t!]
\centering
\includegraphics[width=8.3cm]{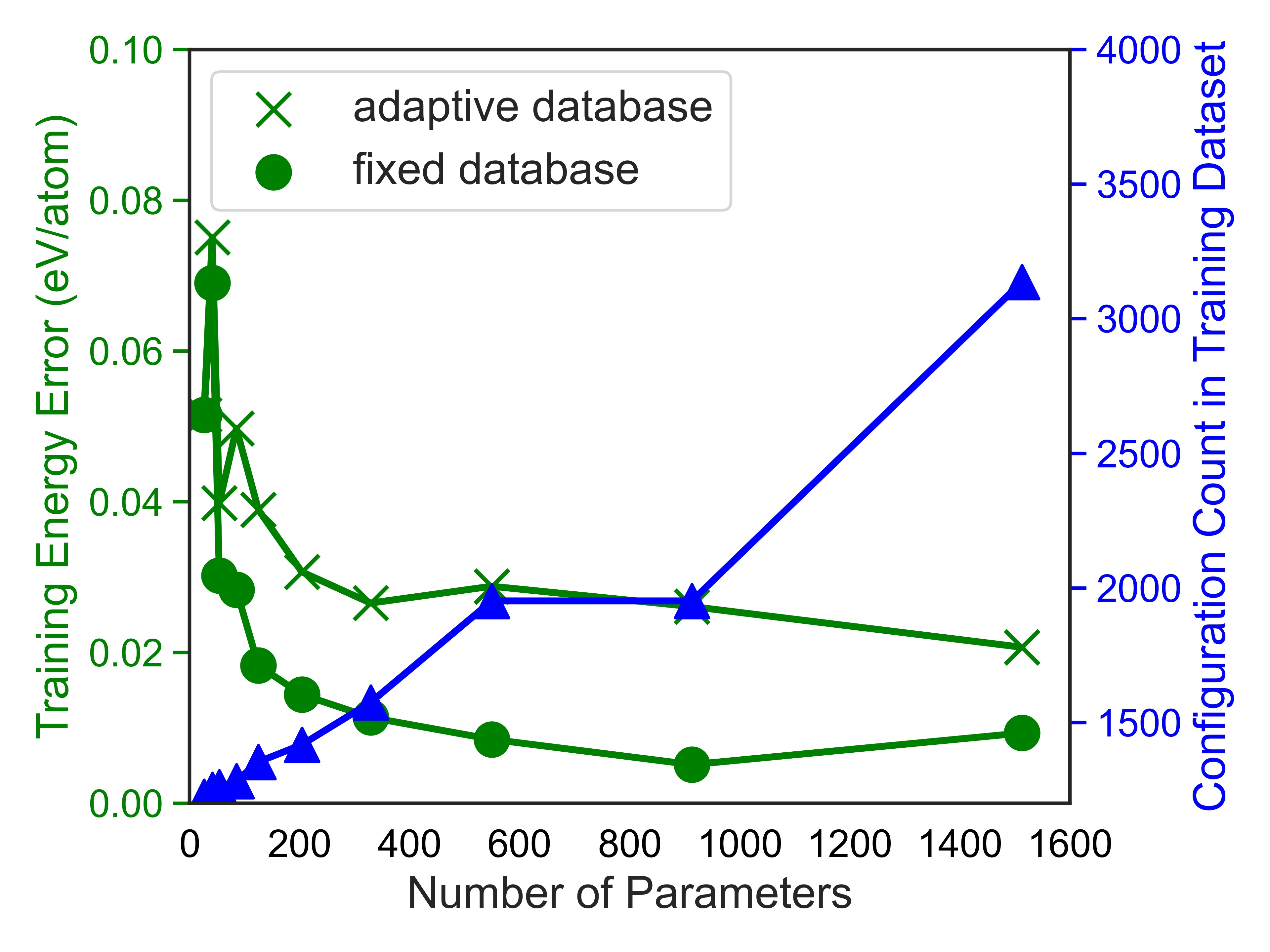}
\caption{The blue data points indicate how the number of configurations in the training dataset increases in tandem with the model complexity. Note that no risky configurations were detected after upgrading from 549 model parameters to 913 parameters. This is likely associated to the stochastic nature of the MLIP-2 fitting procedure. The green points indicate that the training error involving fixed-size and adaptive-size databases both decrease as the model complexity increases; training using an adaptive-size database leads to a slightly higher training error compared to training using a fixed-size database. }
\label{fig:2}
\end{figure}

In general, sophisticated physical properties in atomic simulations require complex force field models to provide reliable predictions. Unfortunately, complex models may not generalize well to unobserved configurations. In this study, the generalization capability in MD simulations is improved by introducing an adaptive-size database, although this comes at the cost of a training error gap with the fixed data set. In Figure \ref{fig:2}, the complexity of a force field is characterized by the number of parameters it has, and more complex force fields with more parameters will give lower training errors. Noteworthy, the number of DFT calls increasingly grows after starting our scheme in Figure \ref{fig:1}, but no clear pattern was observed as to which types of configurations were classified as risky. 

\begin{figure}[t!]
\centering
\includegraphics[width=8.5cm]{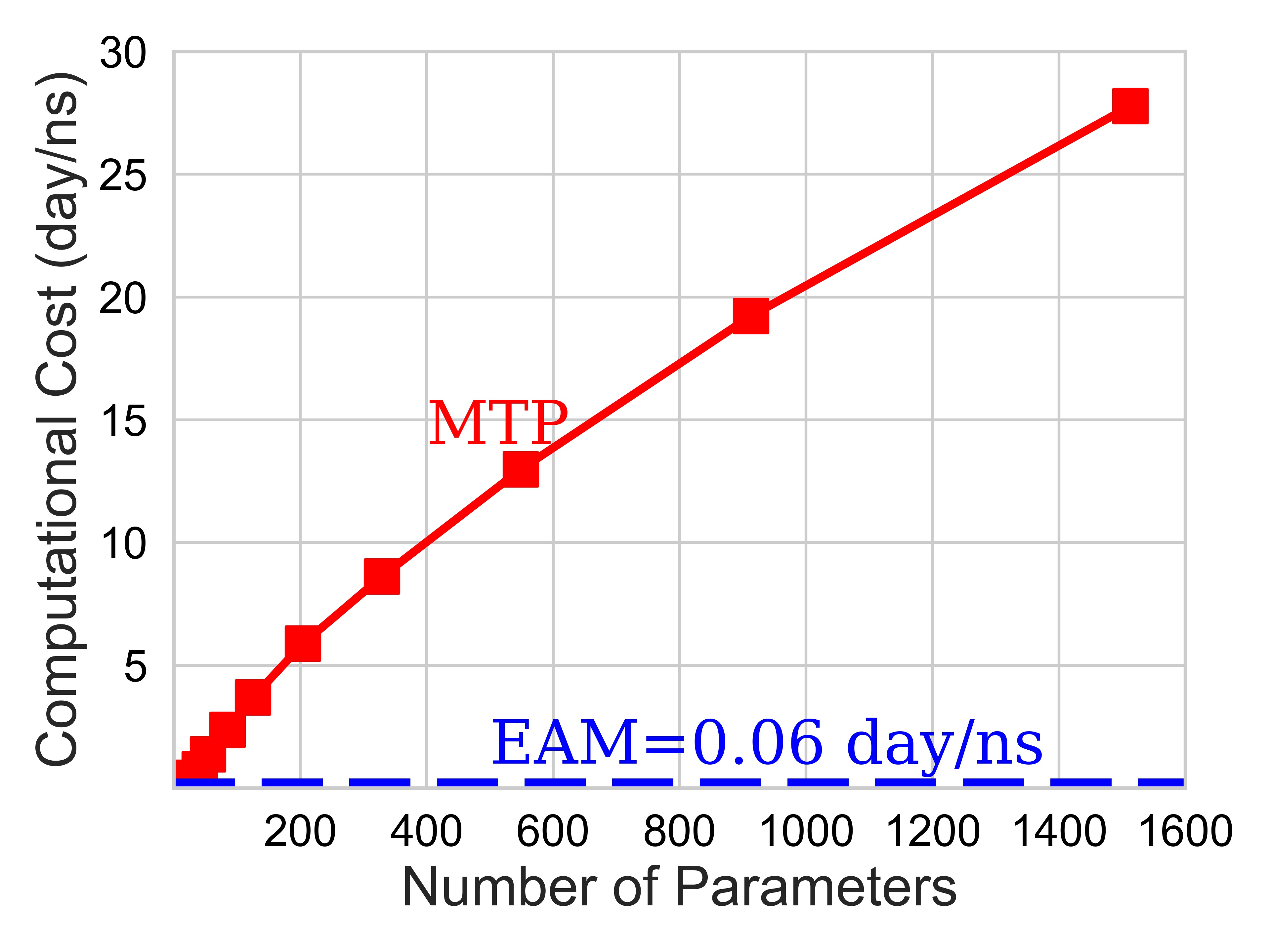}
\caption{The computational cost of 2000-atom MD runs performed on a single computer core increases slightly sub-linearly with the number of model parameters. The 1513-parameter MTP is nearly 500 times more resource-intensive than a typical EAM potential.}
\label{fig:3}
\end{figure}

As illustrated in Figure \ref{fig:3}, EAM models are much more computationally efficiency than MTPs. MTPs with more parameters require more computations to make predictions, which increases their computational burden. The relationship between the number of model parameters and computational cost is slightly sublinear. 

\begin{figure}[t!]
\hspace*{-0.08\linewidth}
\includegraphics[width=9cm]{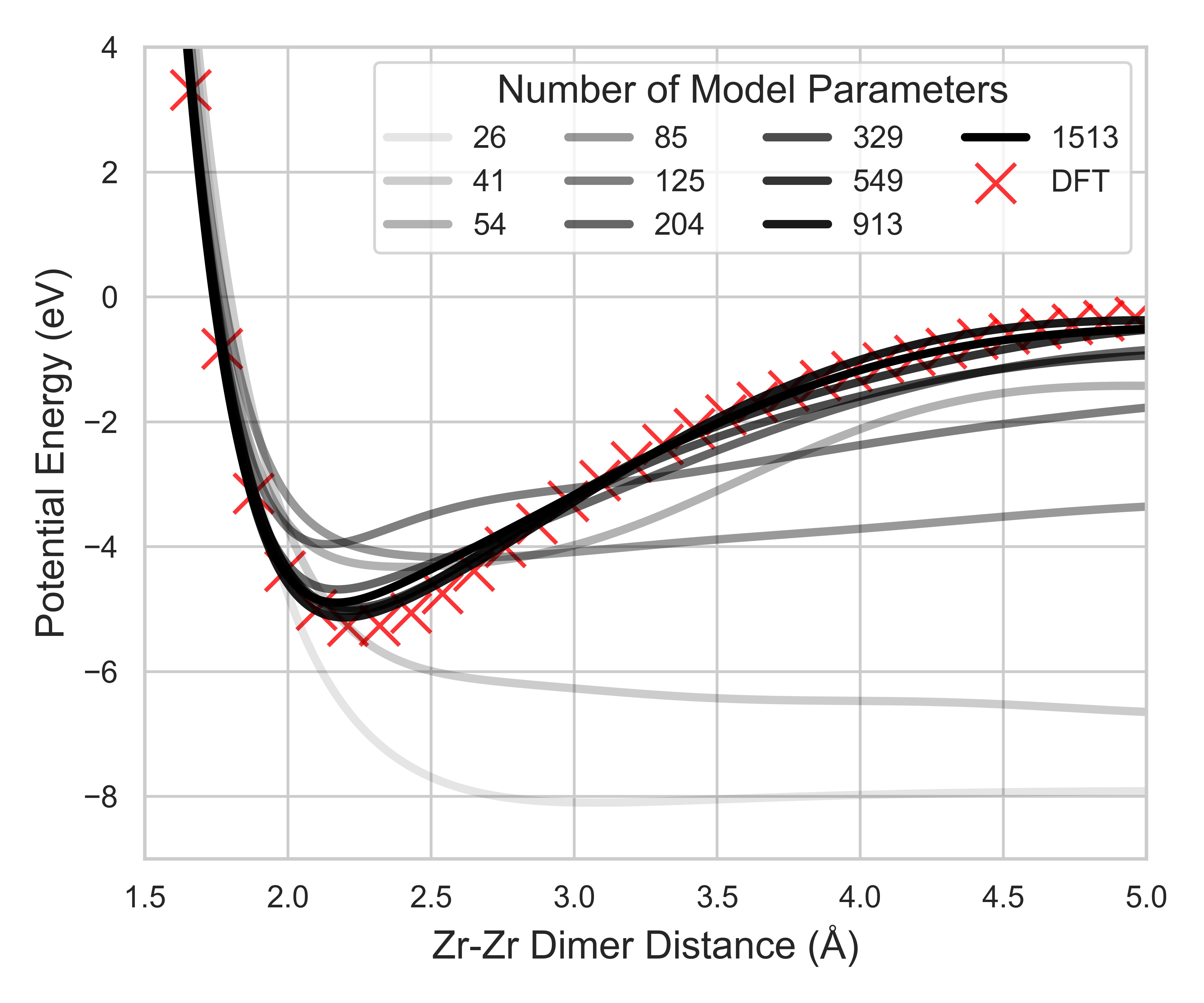}
\caption{Potential energy of Zr-Zr dimer in vacuum as a function of interatomic distance. Note that these Zr-Zr dimers were not explicitly included in the training set. As the number of model parameters increases, the potential energies of Zr-Zr dimers in vacuum converge to the DFT benchmark.}
\label{fig:4}
\end{figure}

As demonstrated in Figure \ref{fig:4}, increasing the MTP complexity leads to an improvement of their ability to capture the Zr-Zr dimer energy. Though these configurations are not directly relevant to realistic solid-state systems, pair-wise interactions are a critical component of many-body interactions. Force fields with fewer model parameters are unable to qualitatively predict the potential energy landscape of the dimer in a vacuum; for instance, many of the low-complexity MTPs do not exhibit a ``well" shape. Despite being one hundred times more computationally demanding than EAM potentials, these low-complexity MTPs perform worse than EAM potentials in this out-of-sample system. 

\begin{figure}
\hspace*{-0.07\linewidth}
\includegraphics[width=9.1cm]{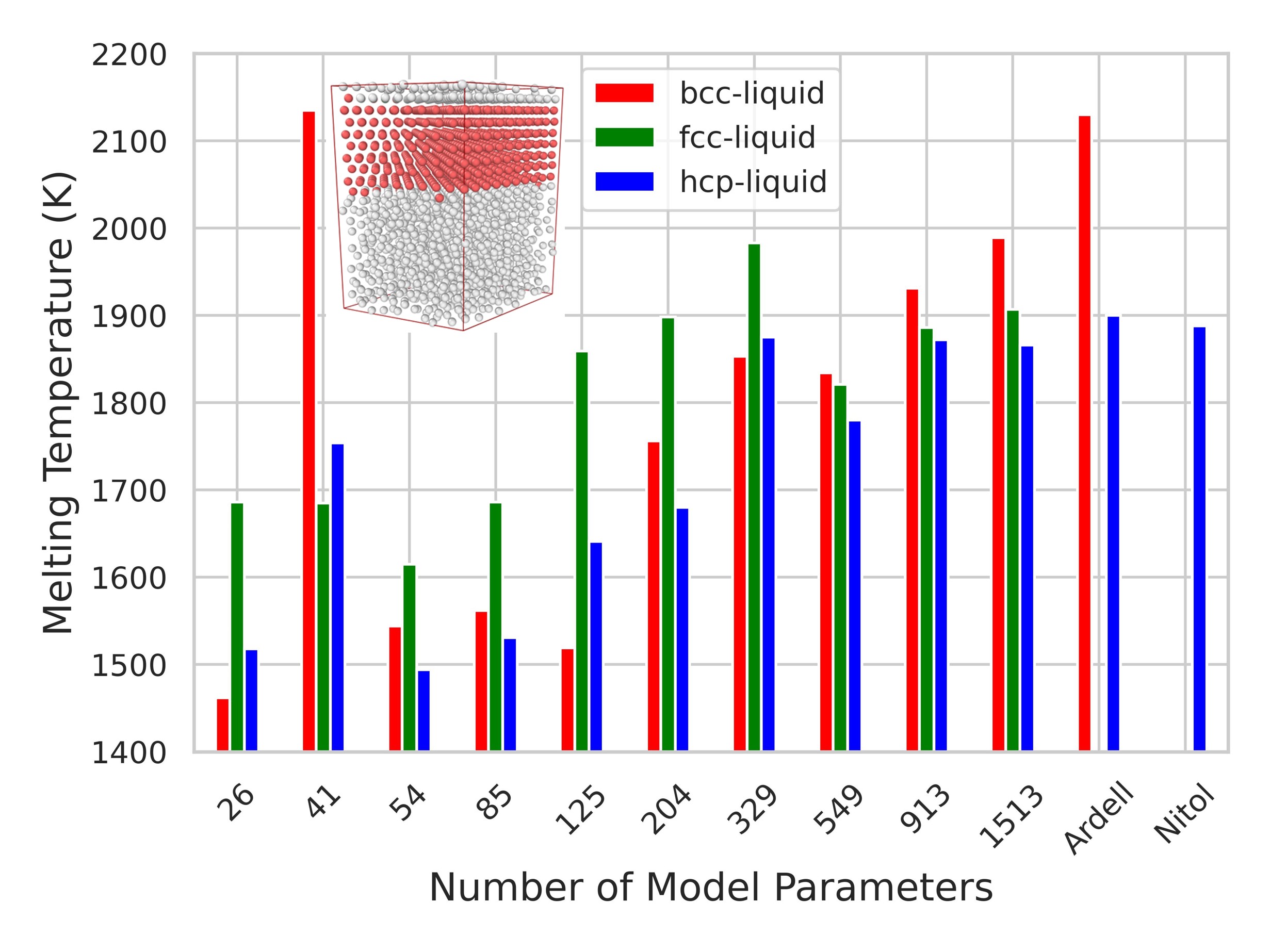}
\caption{Melting point as predicted by the MTP potentials were calculated using the moving interface method. See inset for a typical configuration; red atoms are crystalline, and grey atoms are non-crystalline. The experimental melting temperature, a theoretical “reduced” melting temperature for $\alpha$-Zr\cite{ardell1963calculation} and a melting temperature obtained from a very large neural network potential\cite{nitol2022machine} trained by Nitol \textit{et al.} are included as benchmarks.}
\label{fig:5}
\end{figure}

In Figure \ref{fig:5}, melting points of hcp, bcc, and fcc crystal structures are estimated using the moving interface method\cite{morris1994melting,galvin2021molecular} with varying levels of model complexity, indicating complex interatomic interactions arise during high-temperature scenarios. The melting point of hcp is lower than that of fcc, which was in turn lower than that of bcc. MTP models with 549 or more parameters capture this behavior. The MTP-generated values converge to benchmark values with increasing model complexity. The hcp benchmark represents a reduced melting point (1900 K), which is an analytical melting point estimation for $\alpha$-Zr\cite{ardell1963calculation}. The bcc MTP melting point (1990K) is lower than the experimental melting point (2110K). For reference, the neural network potential presented by Nitol\cite{nitol2022machine}, trained on a DFT dataset containing over 10,000 configurations, led to a bcc melting point of $\sim$1900K. This suggests the $\sim$100K discrepancy between our MTP potential predictions and experimental predictions may be due to the limitations of the underlying DFT calculations. Additionally, good generalization capability in melting behaviors was observed as well, considering that no configurations associated with melting points were directly included in the training set. 

\begin{figure}
\hspace*{-0.07\linewidth}
\includegraphics[width=9cm]{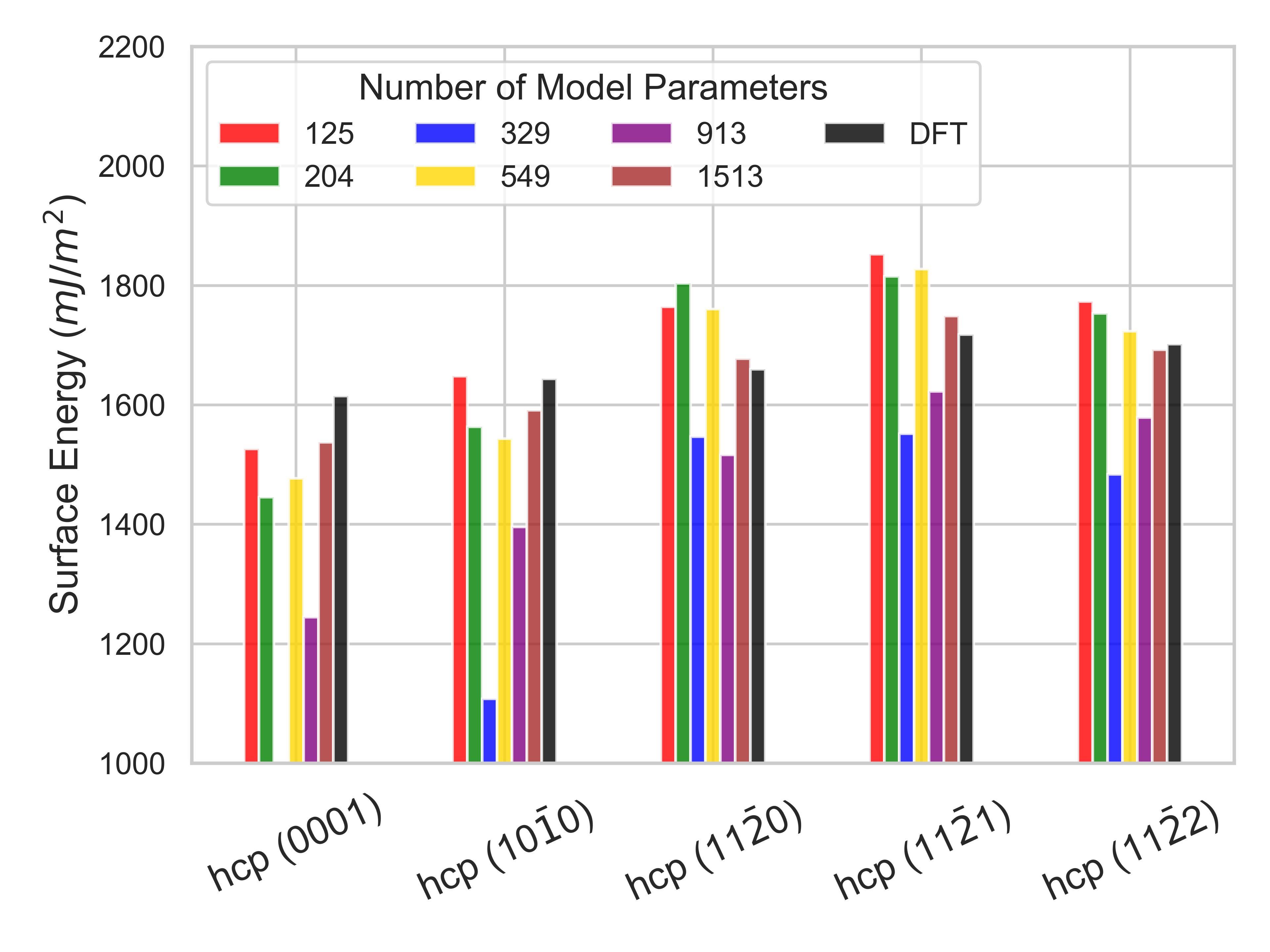}
\caption{Surface energies of $\alpha$ Zr as predicted by our MTPs possessing 125 parameters or more. Surface energies of $\alpha$ Zr gradually converge to the DFT benchmark.}
\label{fig:6}
\end{figure}

Atomic interactions in surface structures refer to complex many-body scenarios with highly asymmetric atomic surroundings. As displayed in Figure \ref{fig:6}, ``stable" surface energies of zirconium on hcp (0001), hcp (10$\bar{1}$0), hcp (11$\bar{2}$0), hcp (11$\bar{2}$1), and hcp (11$\bar{2}$2) are discovered only starting from 125 to 1513 parameters. In contrast, "Unstable" surface energies are observed in MTPs with less than 125 parameters because their surface energies are found to be sensitive to changes in system size, while predictions from complex MTPs are more robust. Similar phenomena were also noted in reference to linear machine learning force fields\cite{grigorev2023calculation}. Noticeable deviations from DFT benchmarks are observed, but their deviations gradually decrease as the model complexity increases. 

Again, the ranking of surface energies for these surfaces is correctly predicted with complex MTPs, agreeing with the DFT benchmark. These suggest that using more sophisticated force field models can improve the accuracy and scalability of system size in surface energy predictions. Once again, the low-complexity MTPs perform worse than EAM potentials (which do not suffer from unstable surface energies), despite their higher computational costs.

\begin{figure*}[t!]
\centering
\includegraphics[width=16cm]{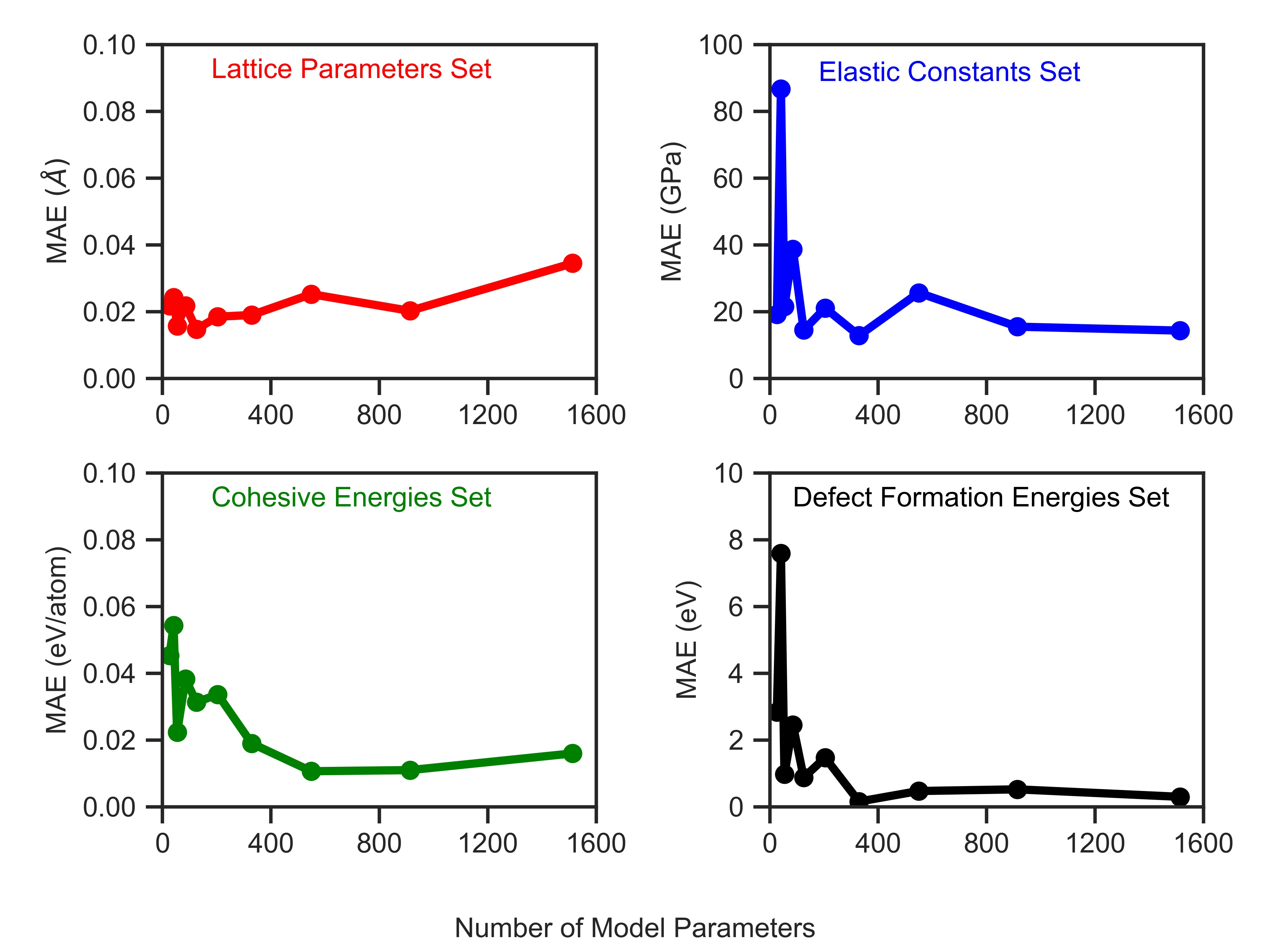}
\caption{List of mean absolute error (MAE) calculating the average absolute difference between physical properties (lattice parameters, elastic constants, cohesive energies, and point defect formation energies) and their corresponding DFT benchmarks in the context of multiple model complexity. A table containing all of these values are included in the Supplementary Information. }
\label{fig:7}
\end{figure*}

A broad range of physical parameters are typically reported when new interatomic force-fiels are introduced, related to lattice parameters, elastic constants, cohesive energies and defect formation energies. In order to study how these values evolve as a function of model complexity, we constructed four sets (lattice parameters, elastic constants, cohesive energies, and defect formation energies) to facilitate analysis. More details about the content of each set is provided in the Supplementary Information. \ref{fig:7} shows mean absolute error (MAE) for these sets of physical features as a function of model complexity. 

The lattice parameters set contains the 4 lattice parameters of Zr crystals (hcp, fcc, and bcc). All MTPs have an MAE less than 0.04 \AA\ in lattice parameters, which is an indication that lattice parameters can be easily predicted using MTPs. Nonetheless, when the model complexity is varied, small volatility in lattice parameters is detected. 

In contrast to lattice parameters, a large MAE in elastic constants set is revealed even with complex models; this set includes 6 hcp elastic constants, 3 fcc elastic constants, and 3 bcc elastic constants. Elastic constants are more sensitive to minute changes in the energy landscape since they are the second derivative of energies with respect to applied strains, leading to a larger MAE. Nevertheless, MAE in elastic constants set consistently decreases with increasing model complexity.

The cohesive energies set (3 cohesive energies for hcp, fcc, and bcc, separately) and formation energies set (4 point defect formation energies for vacancies, di-vacancies, basal octahedrals) exhibit the same trend as the elastic constants set but with a much lower MAE. Overall, discernible features can be easily captured by MLFFs, while capturing subtler feature changes or indirect features is challenging. A solution would be to explicitly target these features and give them a large weight in the cost function, as, e.g., in Ref. \cite{grigorev2023calculation}. In other words, the learning likely requires increased supervision.  
 

Training the MTPs is a global optimization problem. The global optimization process is sensitive to the choice of pre-optimization parameters; that is, the local minima which will be reached depends on the random seeds used for the optimization. In our study, 100 trials with 1513 parameters are conducted to estimate how sensitive physical properties are to change of random seeds.

\begin{figure}[t!]
  \centering
  \begin{subfigure}{1\linewidth}
    \centering
    \includegraphics[width=\linewidth]{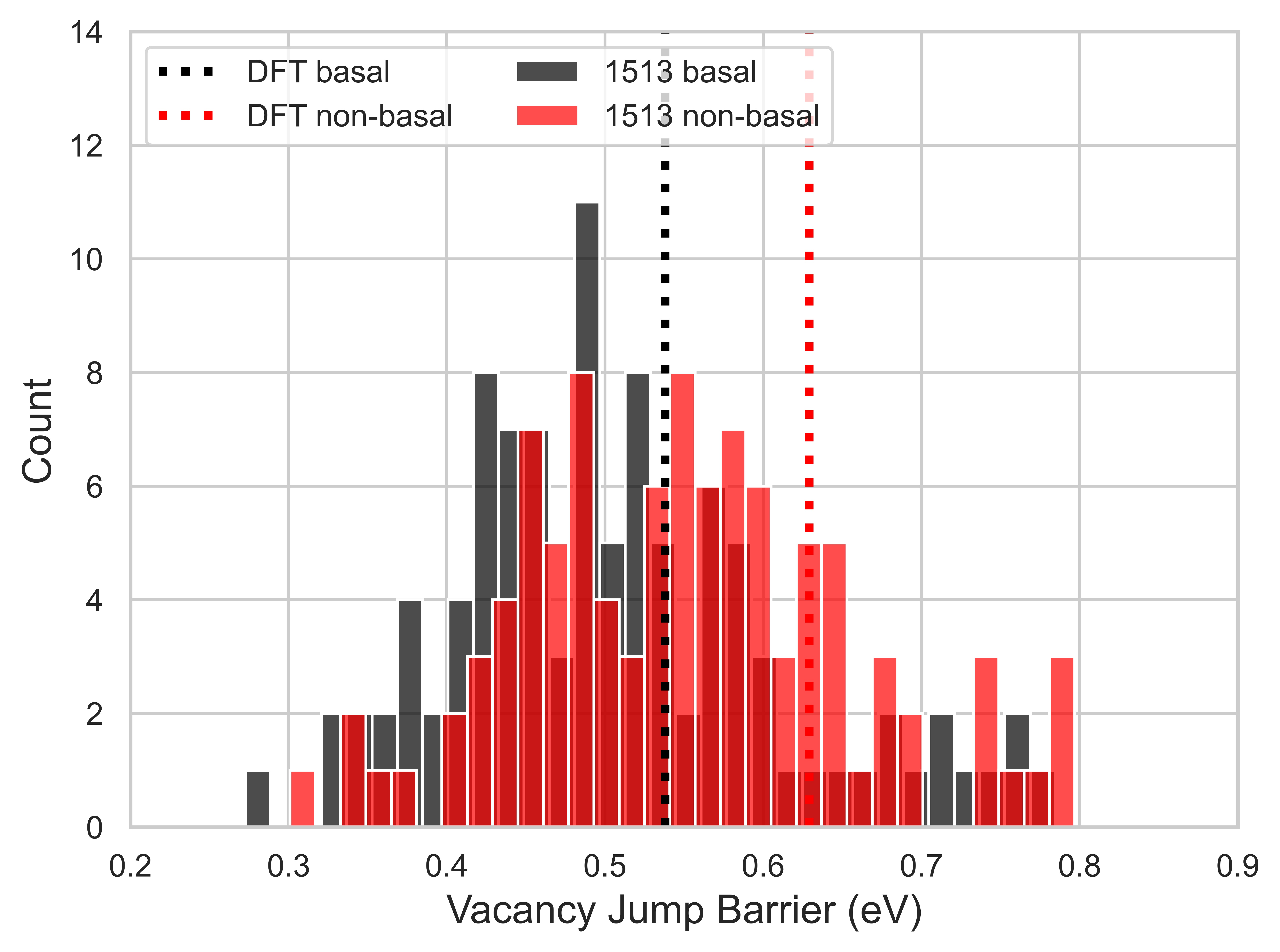}
    \label{fig:9a}
  \end{subfigure}
  \begin{subfigure}{1\linewidth}
    \centering
    \includegraphics[width=\linewidth]{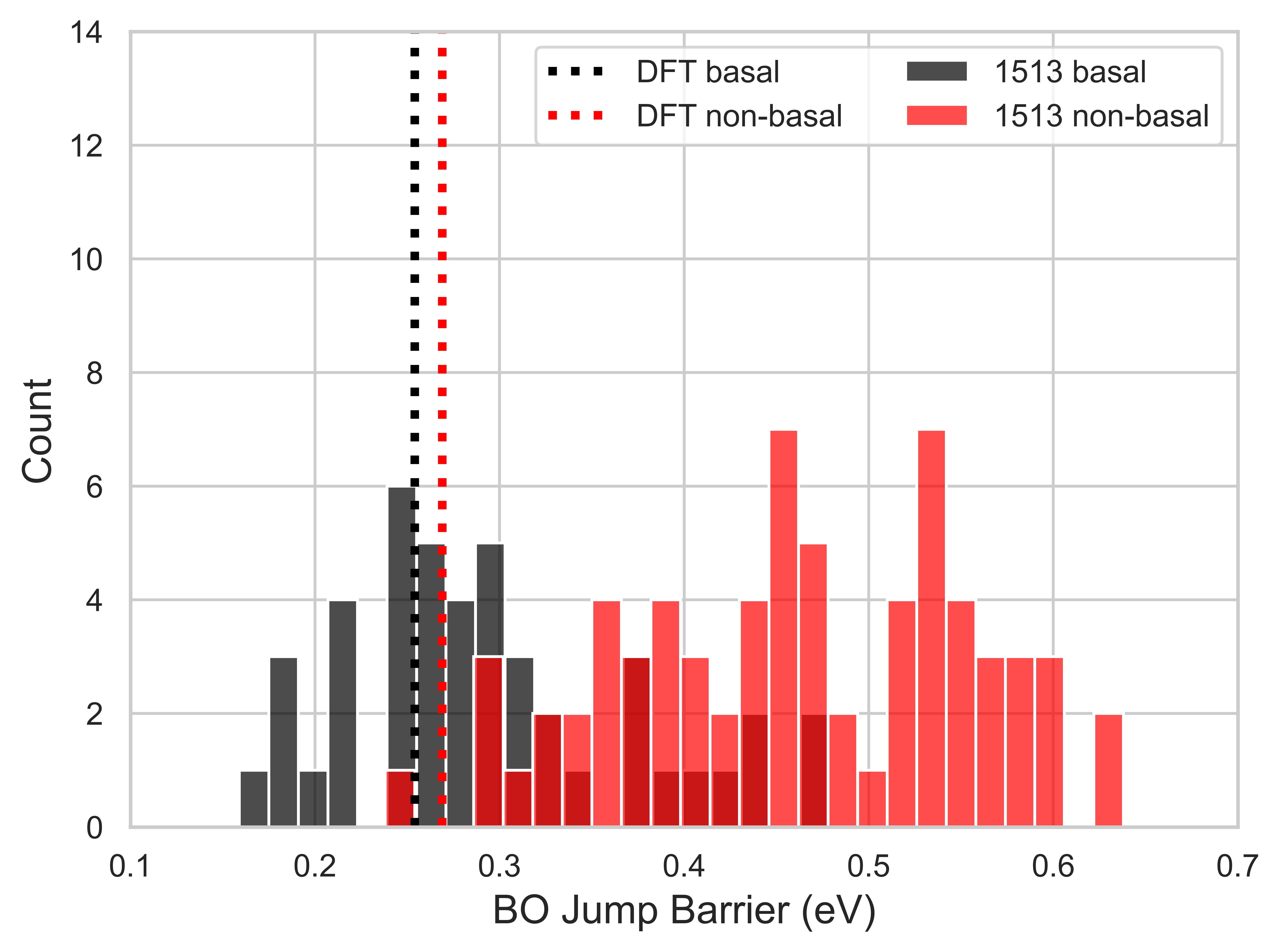}
    \label{fig:9b}
  \end{subfigure}
  \caption{Distribution of migration barriers of vacancies and SIAs parallel and perpendicular to the basal planes in $\alpha$-Zr as predicted by 100 independently optimized 1513-parameter MTPs using the same training dataset. Migration barriers were calculated using the nudged elastic band method. DFT benchmarks are represented with a dotted line.}
  \label{fig:8}
\end{figure}

Predicting migration paths can be challenging because the paths themselves are potential-dependent and are therefore not directly included during force field fitting. Also, system size can directly affect migration barriers and optimal migration paths. As can be seen in Figure \ref{fig:8}, the distribution of vacancy jump barriers is normally distributed; randomness significantly affects vacancy migration barriers. This uncertainty induced by random initialization has an important impact on the MTPs' ability to capture small features or indirect features. All 100 MTPs with 1513 parameters predict stable, first-order vacancy migration paths. Statistically, the in-basal-plane vacancy jumps have a lower energy barrier than out-of-basal-plane jumps, in agreement with DFT. Here, 87 out of 100 MLFFs capture this vacancy anisotropic phenomenon correctly. 

\begin{figure*}[t!]
\centering
\includegraphics[width=18.2cm]{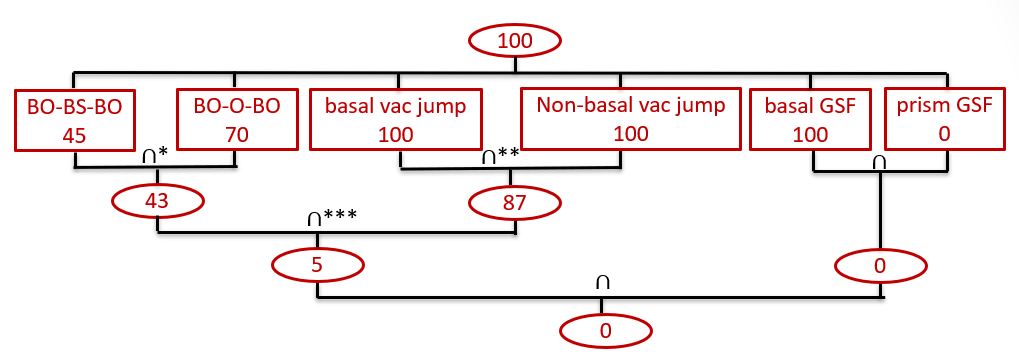}
\caption{One hundred independent 1513-parameter MTPs were optimized using different initial parameters using the same training database. The tree indicates how many of these MTPs capture physical properties related to anisotropy of point defect diffusion and plasticity. $\bigcap^*$ indicates 43 MTPs are consistent with DFT in that the jump barrier for BO-BS-BO SIA migration (in-plane) is lower than that of BO-O-BO SIA migration (out-of-plane). $\bigcap^{**}$ indicates there are 87 MTPs consistent with DFT in that in-plane vacancy jump barriers of vacancy are lower than out-of-plane jump barriers. $\bigcap^{***}$ indicates 5 MTPs can correctly predict the order of both SIA and vacancy in-plane and out-of-plane jump barriers. Out of the 100 1513-parameter MTPs generated, none were associated with both qualitatively correct prism I $\gamma$-surfaces.}
\label{fig:9}
\end{figure*}

In Zr, the basal octahedral (BO) SIA is the most stable SIA configuration. Not all 100 MTPs capture this feature; the octahedral (O) site was the most stable according to 6 MTPs in contradiction with reported DFT results\cite{samolyuk2014analysis,peng2012stability}. As represented in Figure \ref{fig:9}, instead of normal distribution, the distributions of BO migration barriers are skewed for basal and non-basal directions. Statistically, migration paths BO-BS-BO display a lower migration barrier than BO-O-BO, emphasizing BO sites preferentially migrate along the basal direction in Zr\cite{woo1988theory}. As displayed in Figure \ref{fig:9}, of the total 100 MTPs, 43 models capture this SIA anisotropic phenomenon correctly. Furthermore, a good agreement between the expected value of the BO jump barrier distribution along the basal direction and the DFT benchmark is discovered. In contrast, force field models predict a higher expected value of BO jump barrier along the non-basal direction compared to reported DFT results. 

We also characterized the basal and prismatic GSFE surfaces (see Supplementary Information to visualize these GSFE surfaces). As summarised in Figure \ref{fig:9}, none of the 100 MTPs exhibit a spurious local minima along the [1$\overline{1}$00] in the basal plane. In other words, this artifact which frequently leads to dislocation dissociation when using EAM potentials is systematically avoided by all of these high-complexity MTPs. On prismatic GSFE surfaces, DFT predicts the presence of a local minimum along the [1$\overline{2}$10] direction \cite{clouet2012screw},  which is associated with a 1/3$<$1$\overline{2}$10$>$\{10$\overline{1}$0\} dislocation dissociation into two partial dislocations. This minimum is absent from EAM potentials (it is typically found on a nearby location of the prismatic GSFE surface, leading to dislocation partials which are not perfectly aligned with the glide plane). In contrast, none of the 100 MTPs considered here predicted a humped prismatic GSFE curve along [1$\overline{2}$10] direction rather than a bell-shaped curve. 

Furthermore, we generated another randomly initialized 100 MLFFs (1513 parameters each) using generated MTP-26 training dataset and introduced additional prismatic GSFE structures. Despite these efforts, no significant improvements in the prismatic GSFE curve were observed (see Supplementary Information). To verify if this is an intrinsic limitation of the MTP formalism, we trained an MTP using an offline approach, with a heavy emphasis on the GSFEs. By doing so, we successfully trained an MTP capable of capturing the humped prismatic GSFE curve along [1$\overline{2}$10] direction. Unfortunately, we observed some non-physical behavior when testing this potential, including nonphysical dimer potential energies, the presence of negative values for prismatic generalized stacking fault energies, data underflow issues when performing MD runs involving point defect migration, etc.; therefore we do not recommend its use. More details about this MTP trained offline are available as Supplemental Information.

\section{Conclusion}
To summarize, a hybrid scheme involving small-cell offline and active learning is proposed and applied to Zr. The main conclusions are as follows:

1. Although high-complexity MTPs are capable of capturing subtle, indirect, or implicit physical properties, EAM and other physics-based potentials are likely preferable over MTPs possessing fewer than $\sim$500 parameters. Unlike low-complexity MTPs, EAM-type potentials do not display severe nonphysical behavior, while being much faster. 

2. The aforementioned nonphysical behavior mostly occurred when calculating defect formation energies in large simulation boxes. Surprisingly, these issues plaguing large-sized configurations were addressed by training over small-sized configurations. The prediction results over various scenarios (simple pair-wise interactions, complicated interactions at high-temperature, etc.) are quite encouraging considering the MTPs trained herein were not specifically trained or designed to reproduce these results. 

3. Statistical uncertainty is unavoidable if non-linear optimization is involved. The distributions we obtained varied depending on the physical properties being evaluated, but they were generally centered around the reference DFT values. Good point defect migrations barriers and GSFE surfaces can be captured by a subset of the most complex (1513-parameter) MTPs we trained. Overall, 5 out of 100 MTPs get both SIA and vacancy migration barriers in the correct order. All 100 MTPs are associated with qualitatively correct basal GSFE surfaces, but none of them can correctly capture prismatic GSFE surfaces.  Hence, we recommend users pick the potential dubbed MTP-PD (point defect) which could be used to simulate either point defect formation and migration respectively. Further details about MTP-PD can be found in the Supplementary Information.   

4. Increasing the model complexity of the MTPs may have resolved some of the remaining descrepencies between DFT and MTPs. However, training even more complex MTPs would have involved important software- and hardware-based practical issues, in addition to adding even more computational cost to the model; we therefore stopped once reaching MTP-level 26. 

5. Our work used the D-optimality criterion as implemented in MLIP-2 \cite{novikov2020mlip} to calculate the reliability of predictions. Other metrics (e.g. Bayesian estimators) may perform better and should be assessed in future work.

6. Our work was limited to the MTP formalism as implemented in MLIP-2 \cite{novikov2020mlip}. Using different descriptors and regression models may lead to superior results. Our proposed iterative hybrid small-cell training scheme can be applied to these other MLFF frameworks.

7. Our work gives a path toward semi-supervised potential training by consistently augmenting adaptive training datasets and model complexity. High-complexity MTPs are required to achieve good generalization capability while maintaining prediction accuracy. It is also worth noting that high-complexity MLFFs (our MTPs or neural network potentials\cite{nitol2022machine}) require sufficient third-party validations on various physical properties of zirconium to ensure their accuracy and transferability.

8. All possible local atomic environments are not guaranteed to be completely captured by our framework, in part because our approach necessitates a certain degree of supervision to choose initial configurations for small-cell testing. For MLFFs training to be fully automated, methods to generate training sets encompassing a large proportion of the feature space are necessary. The entropy-maximization approach developed by Karabin \textit{et al.} is one such promising approach \cite{karabin2020entropy}.

\begin{acknowledgement}
The authors thank the Digital Research Alliance of Canada (formerly known as Compute Canada) for the generous allocation of computing resources. The research was supported by the Natural Sciences and Engineering Research Council of Canada (NSERC) and the NSERC/UNENE Industrial Research Chair in Nuclear Materials at Queen’s.
\end{acknowledgement}

\begin{suppinfo}
The MTP files and training dataset files are available at the following GitLab repository: \url{https://gitlab.com/yluo13/mtp-zr}.
\end{suppinfo}

\bibliography{ref}
\end{document}